\documentclass[twocolumn,aps,pra]{revtex4}
\usepackage[usenames,dvipsnames]{xcolor} % only for the comments
\usepackage{ulem} % only here for strikethrough-font

\usepackage{lineno}
\usepackage{subfigure}
\usepackage{sublabel}
\usepackage{dcolumn}
\usepackage{amsmath,amssymb}
\usepackage{bm}
\usepackage{bbm}
\usepackage{color}
\usepackage{overpic}
\usepackage{latexsym}
\usepackage{epstopdf}
\usepackage{color}
\usepackage[english]{babel}
\usepackage{times}
\usepackage{latexsym}
\usepackage{stmaryrd}
\usepackage{psfrag,graphicx}
\usepackage{epsf}
\usepackage{subfigure}
\usepackage{amsmath}
\usepackage{amssymb}
\usepackage{amsfonts}
\usepackage{natbib}
\usepackage{epstopdf}
\usepackage{appendix}
\usepackage{enumitem}
\definecolor{mygrey}{gray}{0.35}
\definecolor{myblue}{rgb}{0.2,0.2,0.8}
\definecolor{myzard}{cmyk}{0,0,0.05,0}
\definecolor{mywhite}{rgb}{1,1,1}
\definecolor{myred}{rgb}{0.9,0.1,0.}
\usepackage[colorlinks=true,citecolor=myblue,linkcolor=myblue,urlcolor=myblue]{hyperref}

\newcommand{\tr}{\operatorname{\bf{tr}}}

\newcommand{\id}{\mathbbm{1}}

\begin{document}
\title[Short Title]{Quantifying Coherence}

\author{T.\ Baumgratz}
\author{M.\ Cramer}
\author{M.B.\ Plenio}
\affiliation{Institut f\"{u}r Theoretische Physik, Albert-Einstein-Allee 11,
Universit\"{a}t Ulm, 89069 Ulm, Germany}

\begin{abstract}
We introduce a rigorous framework for the quantification of coherence and identify intuitive and easily computable measures of coherence. We achieve this by adopting the viewpoint of coherence as a physical resource. By determining defining conditions for measures of coherence we identify classes of  functionals that satisfy these conditions and other, at first glance natural quantities, that do not qualify as coherence measures. We conclude with an outline of the questions that remain to be answered to complete the theory of coherence as a resource.
\end{abstract}

\maketitle

{\it Introduction.---}Coherence, being at the heart of interference phenomena, plays a central role in physics as it enables applications that are impossible within classical mechanics or ray optics. The rise of quantum mechanics as a unified picture of waves and particles further strengthened the prominent role of coherence in physics. Indeed, in conjunction with energy quantization and the tensor product structure of state space, coherence underlies phenomena such as multi-particle interference and entanglement that play a central role in applications of quantum physics and quantum information science.

Quantum optical methods provide an important set of tools for the manipulation of coherence, and indeed, at its basis lies the formulation of the quantum theory of coherence \cite{Glauber63,Sudarshan63}. Here, coherence is studied in terms of phase space distributions and multi-point correlation functions to provide a framework that relates closely to classical electromagnetic phenomena. While this is helpful in drawing intuition from classical wave mechanics and identifies those aspects for which quantum coherence deviates from classical coherence phenomena, it does not provide a rigorous and unambiguous framework. The development of such a quantitative framework for coherence gains further urgency in the light of recent discussions concerning the role of coherence in biological systems~\cite{HuelgaP13} which can benefit from a more rigorous approach to the quantification of coherence properties.

The development of quantum information science over the last two decades has led to a reassessment of quantum physical phenomena such as entanglement, elevating them from mere tools to ``subtly humiliate the opponents of quantum mechanics'' \cite{Bennett} to {\it resources} that may be exploited to achieve tasks that are not possible within the realm of classical physics. This viewpoint, then, motivates the development of a quantitative theory that captures this resource character in a mathematically rigorous fashion. The formulation of such resource theories was initially pursued with the quantitative theory of entanglement \cite{vedral1998entanglement,plenio2007introduction} which led to the view that constraints [e.g., the restriction to local operations and classical communication (LOCC)] that prevent certain physical operations to be realized define resources (e.g., entangled states) that help to overcome the imposed constraints \cite{BrandaoP08,BrandaoP10}. This viewpoint has proven fruitful not only for the development of applications, but also in providing the impetus for theory to establish a unified and rigorously defined framework for a quantitative theory of physical resources by addressing the three principal issues: (i) the characterization, (ii) the quantification, and (iii) the manipulation of quantum states under the imposed constraints \cite{devetak2008,horodecki2013}. 
This framework is being explored for entanglement~\cite{vedral1998entanglement,plenio2007introduction}, thermodynamics \cite{BrandaoHO+11,GourMN+13}, and reference frames \cite{GourS08} and has led to the recognition of deep interrelations between the theories of entanglement and the second law \cite{BrandaoP08,BrandaoP10}. 

In contrast, a wide variety of measures of coherence is in use (often functions of a density matrix' off-diagonal entries) whose use tends to be justified principally on the grounds of physical intuition. Here, we put such measures on  a sound footing by establishing a quantitative theory of coherence as a resource following the approach that has been established for entanglement in Refs. \cite{vedral1998entanglement,BrandaoP08,BrandaoP10} and for reference frames in Ref.~\cite{GourS08}. We present the basic assumptions of our approach and use these to identify various quantitative and easy-to-compute valid measures of coherence while rejecting others. 

{\it Results.---}At the heart of our discussion lies the characterization of incoherent states together with the notion of incoherent operations, i.e., quantum operations that map the set of incoherent states onto itself. We distinguish between quantum operations with and without sub-selection. These technical definitions lead to an operationally well-defined maximally coherent state which may serve as a unit for coherence.
We collect a set of conditions any proper measure of coherence should satisfy. Prime among them is the requirement of monotonicity under incoherent operations. We, then, discuss several examples---some of which take the form of easy to evaluate analytical expressions. For instance, we find that the {\it relative entropy of coherence}
\begin{equation}
    C_{\text{rel.\ ent.}}(\hat{\varrho})=S(\hat{\varrho}_{\text{diag}})-S(\hat{\varrho}),
\end{equation}
where $S$ is the von Neumann entropy and $\hat{\varrho}_{\text{diag}}$ denotes the state obtained from $\hat{\varrho}$ by deleting all off-diagonal elements,
 and the intuitive {\it $l_1$-norm of coherence}
\begin{equation}
    C_{l_1}(\hat{\varrho})=\sum_{\substack{i,j\\ i\ne j}}|\varrho_{i,j}|,
\end{equation}
are both proper measures of coherence.
In contrast, we find that the sum of the squared absolute values of all off-diagonal elements violates monotonicity.

{\it Incoherent states.---}The first step to defining a coherence measure is to agree which states are incoherent. A natural definition is to fix a particular basis, $\{|i\rangle\}_{i=1,\ldots,d}$, of the $d$-dimensional Hilbert space $\mathcal{H}$ in which we consider our quantum states \cite{remark5}. We call all density matrices that are diagonal in this basis {\it incoherent} and, henceforth, label this set of quantum states by $\mathcal{I}\subset \mathcal{H}$ \cite{remark4}. 
Hence, all density operators $\hat{\delta}\in\mathcal{I}$ are of the form
\begin{equation}
\hat{\delta} = \sum_{i=1}^{d} \delta_{i} |i\rangle\langle i|.
\end{equation}

{\it Incoherent operations.---}The definition of coherence monotones (and, thus, coherence quantifiers) requires the definition of operations that are incoherent---just as in entanglement theory the definition of entanglement monotones requires a definition of non-entangling operations. There, this definition is determined by practical considerations, namely locality constraints, which leads to the definition of LOCC operations. Here, we characterize the set of incoherent physical operations as follows. Quantum operations are specified by a set of {\it Kraus operators} $\{\hat{K}_{n}\}$ satisfying $\sum_{n}\hat{K}_{n}^{\dagger}\hat{K}_{n}=\id$. We require the incoherent operators to fulfil $\hat{K}_{n}\mathcal{I}\hat{K}_{n}^{\dagger}\subset\mathcal{I}$ for all $n$ \cite{remark1}. This definition guarantees that in an overall quantum operation $\hat{\varrho}\mapsto \sum_n\hat{K}_{n}\hat{\varrho}\hat{K}_{n}^{\dagger}$, even if one does not have access to individual outcomes $n$, no observer (e.g., one who {\it does} have access to these outcomes) would conclude that coherence has been generated from an incoherent state. Hence, we do not allow, not even probabilistically, that in any of the arms of the quantum operation coherence is generated from incoherent input states.

We distinguish two classes of quantum operations. (A) The incoherent completely positive and trace preserving quantum operations $\Phi_{\text{ICPTP}}$, which act as
    $\Phi_{\text{ICPTP}}(\hat{\varrho})=\sum_n\hat{K}_n\hat{\varrho}\hat{K}_n^\dagger$,
where the Kraus operators $\hat{K}_n$ are all of the same dimension $d_{\text{out}}\times d_{\text{in}}$ and satisfy $\hat{K}_{n}\mathcal{I}\hat{K}_{n}^{\dagger}\subset\mathcal{I}$.
Note that this formulation of quantum operations implies the loss of information about the measurement outcome which may, however, be available in principle.

This leads us to the second class of operations. (B) Quantum operations for which measurement outcomes are retained (depending on the context, called {\it measuring}, {\it selective} or {\it stochastic} operations) and, therefore, permit sub-selection according to these measurement outcomes. These are also defined by Kraus operators $\hat{K}_n$ with $\sum_n\hat{K}^\dagger_n\hat{K}_n=\id$, which now, however, may each have a different output dimension ($\hat{K}_n$ is a $d_{n}\times d_{\text{in}}$ matrix) and are again required to fulfil $\hat{K}_n\mathcal{I}\hat{K}_n^\dagger\subset\mathcal{I}$ for each $n$. Retaining the knowledge of outcomes of the measurement, the state corresponding to outcome $n$ is given by
    $\hat{\varrho}_n={\hat{K}_n\hat{\varrho}\hat{K}_n^\dagger}/{p_n}$
and occurs with probability $p_n=\text{tr}[\hat{K}_n\hat{\varrho}\hat{K}_n^\dagger]$. 

Incoherent Kraus operators that are of particular importance for decoherence mechanisms of single qubits are, e.g., the ones that define the depolarizing, the phase-damping and the amplitude-damping channels \cite{avalle2014,preskillnotes}. Moreover, permutations of modes of dual-rail qubits in linear optics experiments are examples of incoherent operations. With this, we set the framework for a resource theory for quantum coherence. All that follows is deduced from these physically well-motivated definitions.

{\it Maximally coherent state.---}We start by identifying a $d$-dimensional maximally coherent state as a state that allows for the {\it deterministic} generation of {\it all} other $d$-dimensional quantum states by means of incoherent operations. Note that this definition (i) is independent of a specific measure for the coherence and (ii) allows us to identify a unit for coherence to which all measures may be normalized. A maximally coherent state is given by
\begin{equation}
    |\Psi_d\rangle:=\frac{1}{\sqrt{d}}\sum_{i=1}^d|i\rangle,
\end{equation}
because by means of incoherent operations [of type (A) or (B)] alone, any $d\times d$ state $\hat{\varrho}$ may be prepared from $|\Psi_{d}\rangle$ with certainty. 
We show this by explicitly constructing an incoherent operation that
achieves the transformation in Appendix \ref{sec:TheMaximallyCoherentState}.

Two natural questions arise immediately. First, is this maximally coherent state a resource which, when consumed, allows for the generation of all other coherent operations by means of incoherent operations? We demonstrate in Appendix~\ref{sec:Gate} that this is, indeed, the case: Provided with $|\Psi_2\rangle$, every unitary operation on a qubit may be implemented by incoherent operations. Second, one may ask whether incoherent operations introduce an order on the set of quantum states, i.e., whether, given two states $\hat{\varrho}$ and $\hat{\sigma}$, either $\hat{\varrho}$ can be transformed into $\hat{\sigma}$ or vice versa. 
We have to leave this as an open question, but report small progress in Appendix \ref{sec:FiniteCopyTransformations}, for which we note the analogy to the single copy conversion protocol for entangled pure states presented in \cite{vidal1999entanglement,JonathanP99}.

{\it Coherence measures.---}We now collect defining properties that any functional $C$ mapping states to the non-negative real numbers should satisfy in order for it to be a proper coherence measure. First of all, we demand that it vanishes on the set of incoherent states.
\begin{enumerate}[leftmargin=1cm]
    \item[(C1)] $C(\hat{\delta})=0$ for all $\hat{\delta}\in\mathcal{I}$,
\end{enumerate}
i.e., it should be zero for all incoherent states. One may also think of requiring a stronger condition,
\begin{enumerate}[leftmargin=1cm]
    \item[(C1')] $C(\hat{\delta})=0$ iff $\hat{\delta}\in\mathcal{I}$,
\end{enumerate}
which implies non-zero $C(\hat{\varrho})$ whenever $\hat{\varrho}$ contains coherence. Obviously, (C1') implies (C1). Crucially, any proper coherence measure should not increase under incoherent operations of type (A) or (B): 
\begin{enumerate}[leftmargin=1cm]
    \item[(C2a)] Monotonicity under incoherent completely positive and trace preserving maps: $C(\hat{\varrho})\ge C\bigl(\Phi_{\text{ICPTP}}(\hat{\varrho})\bigr)$ for all $\Phi_{\text{ICPTP}}$.
\end{enumerate}
Recall that this ignores the possibility of sub-selection based on measurement outcomes. Retaining measurement outcomes leads to:
\begin{enumerate}[leftmargin=1cm]
\item[(C2b)] Monotonicity under selective measurements on average: $C(\hat{\varrho})\ge \sum_np_nC(\hat{\varrho}_n)$ for all $\{\hat{K}_n\}$ with $\sum_n\hat{K}_n^\dagger\hat{K}_n=\id$ and $\hat{K}_n\mathcal{I}\hat{K}_n^\dagger\subset\mathcal{I}$.
\end{enumerate}

It should be noted that, besides the requirement of monotony under operations of type (A) and (B), one may argue that sub-selection based on measurement outcomes is described by adding a classical flag to the relevant quantum states $\hat{\varrho}_n$, which introduces a third monotonicity requirement. We further comment on this in Appendix \ref{sec:L1AndRelEntSatisfyC2c}, where we show that the relative entropy of coherence and the $l_1$-norm of coherence are also monotonic under these operations, further strengthening their pivotal role. 

Ideally, one would like to identify measures that fulfil both conditions (C2a) and (C2b). We would consider monotonicity under (C2b) more important as it allows for sub-selection, a process available in well controlled quantum experiments. We will see, however, that (C2b) is often harder to verify while (C2a) is automatically satisfied for a wide class of coherence measures. Moreover, from a physical point of view, one would like to ensure that coherence can only decrease under mixing. This leads to our final condition.
\begin{enumerate}[leftmargin=1cm]
\item[(C3)] Non-increasing under mixing of quantum states (convexity): $\sum_n p_nC(\hat{\varrho}_n)\ge C(\sum_n p_n\hat{\varrho}_n)$ for any set of states $\{\hat{\varrho}_n\}$ and
any $p_n\ge 0$ with $\sum_np_n=1$.
\end{enumerate}

Now, coherence measures that satisfy conditions (C2b) and (C3) imply condition (C2a)---again, highlighting the importance of (C2b). This can be seen as follows:
\begin{equation*}
    C\bigl(\Phi_{\text{ICPTP}}(\hat{\varrho})\bigr) = C\Bigl(\sum_np_n\hat{\varrho}_{n}\Bigr) \stackrel{\text{(C3)}}{\leq} \sum_n p_n C(\hat{\varrho}_{n}) \stackrel{\text{(C2b)}}{\leq} C(\hat{\varrho}).
\end{equation*}
In the following, we study natural candidates for coherence measures. All are based on distance measures.

{\it Distance measures.---}For any distance measure between quantum states $\mathcal{D}$, we may define
candidate coherence measures by
\begin{equation}
    C_{\mathcal{D}}(\hat{\varrho})=\min_{\hat{\delta}\in\mathcal{I}} \mathcal{D}(\hat{\varrho},\hat{\delta}),
\end{equation}
i.e., the minimal distance (in the sense of $\mathcal{D}$) of $\hat{\varrho}$ to the set of incoherent quantum states $\mathcal{I}$. By definition, (C1') is automatically fulfilled for all $\mathcal{D}$ with $\mathcal{D}(\hat{\varrho},\hat{\delta})=0$ iff $\hat{\varrho}=\hat{\delta}$, which holds, e.g., if $\mathcal{D}$ is a metric.

In analogy to the theory of entanglement \cite{VedralPRK97}, we may immediately identify an entire class of distance measures $\mathcal{D}$ for which $C_{\mathcal{D}}$ fulfils (C2a): Whenever $\mathcal{D}$ is contracting under CPTP maps, i.e., such that $\mathcal{D}(\hat{\varrho},\hat{\sigma})\ge \mathcal{D} \bigl(\Phi_{\text{CPTP}}(\hat{\varrho}),\Phi_{\text{CPTP}}(\hat{\sigma})\bigr)$ for any completely positive trace preserving map $\Phi_{\text{CPTP}}$, it induces a functional fulfilling (C2a) as then
\begin{equation}
\begin{split}
    C_{\mathcal{D}}(\hat{\varrho})&=\mathcal{D}(\hat{\varrho},\hat{\delta}_*)
    \ge \mathcal{D}\bigl(\Phi_{\text{ICPTP}}(\hat{\varrho}),\Phi_{\text{ICPTP}}(\hat{\delta}_*)\bigr)\\
    &\ge \min_{\hat{\delta}\in\mathcal{I}}\mathcal{D}\bigl(\Phi_{\text{ICPTP}}(\hat{\varrho}),\hat{\delta}\bigr)=C_{\mathcal{D}}\bigl(\Phi_{\text{ICPTP}}(\hat{\varrho})\bigr),
\end{split}
\end{equation}
where we used that $\Phi_{\text{ICPTP}}(\mathcal{I})\subset\mathcal{I}$ and denoted by $\hat{\delta}_*$ the incoherent state minimizing the distance to $\hat{\varrho}$.

Whenever $\mathcal{D}$ is jointly convex, the induced coherence monotone $C_{\mathcal{D}}$ fulfils condition (C3):
\begin{equation}
\begin{split}
&C_{\mathcal{D}}\Bigl(\sum_np_n\hat{\varrho}_n\Bigr) \le \mathcal{D}\Bigl(\sum_np_n\hat{\varrho}_n,\sum_n p_n\hat{\delta}_n^*\Bigr) \\
&\hspace{1.5cm}\le \sum_np_n\mathcal{D}\bigl(\hat{\varrho}_n,\hat{\delta}_n^*\bigr) =\sum_np_nC_{\mathcal{D}}(\hat{\varrho}_n),
\end{split}
\end{equation}
where, for all $n$, $\hat{\delta}_{n}^{*}$ minimizes the distance to $\hat{\varrho}_{n}$. If $\mathcal{D}(\hat{\varrho},\hat{\delta})=\|\hat{\varrho}-\hat{\delta}\|$ with $\|\cdot\|$ any matrix norm, (C3) is automatically implied by the triangle inequality and absolute homogeneity.

Condition (C2b) seems to be much harder to decide. A good starting point for showing (C2b) would be to check whether $\mathcal{D}$ fulfils the conditions given in Ref.\ \cite{vedral1998entanglement}. We proceed by considering specific examples.

{\it Relative entropy of coherence.---}Consider the quantum relative entropy,
    $S(\hat{\varrho}\|\hat{\sigma})=\text{tr}[\hat{\varrho}\log(\hat{\varrho})]-\text{tr}[\hat{\varrho}\log(\hat{\sigma})]$,
and denote the induced measure by $C_{\text{rel.\ ent.}}$. It clearly fulfils (C1) and also (C1'). Further, it is known that the relative entropy is contracting under CPTP maps  and jointly convex~\cite{Lindblad1975,Ruskai2002}, i.e., $C_{\text{rel.\ ent.}}$ satisfies (C2a) and (C3). It also fulfils (C2b), which can be shown following the approach of \cite{vedral1998entanglement} for general selective measurements (see Appendix \ref{sec:RelativeEntropyOfCoherence}). In addition to fulfilling all our requirements for a coherence measure, $C_{\text{rel.\ ent.}}$ permits a closed form solution, avoiding the minimization: Let $\hat{\delta}=\sum_i\delta_i|i\rangle\langle i|\in\mathcal{I}$ and for given $\hat{\varrho}=\sum_{i,j}\varrho_{i,j}|i\rangle\langle j|$ denote $\hat{\varrho}_{\text{diag}}=\sum_i\varrho_{i,i}|i\rangle\langle i|$. Then, 
    $S(\hat{\varrho}\|\hat{\delta})=S(\hat{\varrho}_{\text{diag}})-S(\hat{\varrho})+S(\hat{\varrho}_{\text{diag}}\|\hat{\delta})$,
and hence,
\begin{equation}
    C_{\text{rel.\ ent.}}(\hat{\varrho})=S(\hat{\varrho}_{\text{diag}})-S(\hat{\varrho}).
\label{eqn:RelEntMeasure}
\end{equation}
Employing this formula, we can easily find the maximum possible value of coherence in a state: For any state $\hat{\varrho}$, one has $C_{\text{rel.\ ent.}}(\hat{\varrho})\le S(\hat{\varrho}_{\text{diag}})\le \log(d)$ and this bound is attained for the maximally coherent state defined above. Note that this relative entropy measure was also considered in similar contexts such as, e.g., to quantify superposition and frameness \cite{Aberg06,Gour09,Horodecki05,Angelo2013,Rodriguez2013,Vaccaro2008}. Notably, monotonicity of $C_{\text{rel. ent.}}$ under (C2a) is a special case of a result of Ref.\ \cite{Aberg06}.

{\it $l_p$-norms.---}A very intuitive quantification of coherence would certainly be related to the off-diagonal elements of the considered quantum state. Therefore, quantifying the coherence by a functional depending on the off-diagonal elements is desirable. A widely used quantifier of coherence is given by
\begin{equation}
    C_{l_1}(\hat{\varrho})=\sum_{\substack{i,j \\ i\ne j}}|\varrho_{i,j}|.
\end{equation}
But is it a proper coherence measure in the sense of (C1)--(C3)? If so, it would constitute another intuitive coherence monotone with an easy closed form. It is the measure induced by the $l_1$ matrix norm \cite{HornJ91},
   $ \mathcal{D}_{l_1}(\hat{\varrho},\hat{\delta})=\|\hat{\varrho}-\hat{\delta}\|_{l_1}= \sum_{i,j}|\varrho_{i,j} - \delta_{i,j}|$,
and as such fulfils (C1') and (C3). What is more, (C2b) can be shown directly (see Appendix \ref{sec:l1NorOfCoherence}) such that (C2a) is implied (see the discussion above). Hence, the $l_1$-norm of coherence, together with the relative entropy of coherence, are the most general coherence monotones established in this
manuscript.

One may now ask whether measures induced by other $l_p$ matrix norms serve as proper coherence monotones as well. For instance, consider the measure induced by the squared Hilbert-Schmidt norm; that is 
\begin{equation}
C_{l_2}(\hat{\varrho})= \min_{\hat{\delta}\in\mathcal{I}} \|\hat{\varrho}-\hat{\delta}\|_{l_2}^{2}=\sum_{\substack{i,j\\ i\ne j}}|\varrho_{i,j}|^2.
\end{equation}
In Appendix~\ref{sec:HSNormIsNoMonotone}, we show that $C_{l_2}$ does {\it not} satisfy (C2b), i.e., that there are incoherent operations of type (B), under which $C_{l_2}$ increases. This shows that care must be taken when quantifying coherence: While $C_{l_2}$ might intuitively seem like a good candidate due to its simple structure related to the off-diagonal elements of the quantum state, it does not constitute a valid coherence monotone.

We discuss other potential candidates such as the measures induced by the fidelity and trace norm in Appendix~\ref{sec:OtherCandidates}. 

{\it Outlook.---}In the preceding, we have provided the foundations for a theory of coherence as a resource as well as first results specifically concerning the quantification of coherence. Completion of this theory is a sizeable task that requires a thorough consideration of the questions of the manipulation, quantification,
and exploitation of coherence under this resource viewpoint.

In this work, we have determined the notion of a maximally coherent state, but we have not yet provided a full theory of the interconversion of coherent states by means of incoherent operations. This has two principal aspects. On the one hand, the setting of single copies of coherent states is of considerable interest from the practical point of view as this is most readily accessible in the laboratory. We expect that a theory can be established that proceeds along analogous developments in entanglement theory.  There, the concept of majorization provided the relevant structure that determined the interconvertibility of states \cite{Nielsen99,vidal1999entanglement,JonathanP99} and enabled the exploration of concepts such as catalysis \cite{JonathanP99b}. Some progress in this direction has been reported in Ref. \cite{Aberg13} for a specific set-up but with a different class of allowed quantum operations. Whether such a phenomenon also occurs for this resource theory of coherence or whether a total order on quantum states can be established needs to be explored. 

On the other hand, the asymptotic limit of infinitely many identically prepared copies of a coherent state and its interconversion by incoherent operations is of interest as it may provide a link to thermodynamical concepts such as the second law \cite{PopescuR96,BrandaoP08,BrandaoP10}, by enabling reversible  interconversion of coherent resources and, by invoking natural continuity requirements such as asymptotic continuity \cite{Vidal00}, it may lead to the identification of a unique coherence measure. The latter we expect to be realized by the relative entropy of coherence in close analogy to the development in entanglement theory~\cite{BrandaoP08,BrandaoP10}. Reference \cite{Rodriguez2013} takes steps in this direction and provides a thermodynamic interpretation of the relative entropy measure of coherence in the context of thermodynamic equilibria for decoherence processes.

A second aspect of the manipulation of coherence concerns its exploitation as a resource when only incoherent operations and a supply of coherent states is available \cite{Eisert2000,Aberg13}. In a first step, we have demonstrated that any (coherent) unitary operation can be realized in this fashion. The resource optimal protocols and the generation of the most general quantum operation from these resources has not yet been established. This in turn motivates questions such as the coherence cost of quantum operations and the dual question of coherence power of operations, again closely mirroring analogous developments in entanglement theory \cite{Eisert2000,ZanardiZF00}.

It is likely that each of the three lines of enquiry above will lead to the definition of sensible and good coherence measures, each of which will be related to the efficiency of certain coherence transformations. This will then provide a well rounded picture of the quantification of coherence as a resource.

All of the considerations above implicitly assumed the finite dimensional setting, but this is neither necessary nor desirable as there are very relevant physical situations that require infinite dimensional systems for their description. Most notable the quantum states of light, that is quantum optics, with its bosonic character requires infinite dimensional systems, harmonic oscillators, for their description. Hence, a quantum theory of coherence in infinite dimensional systems is needed. Again, closely mirroring the development of entanglement theory mathematical problems concerning continuity that are inevitably emerging can be addressed by requiring energy constraints \cite{EisertSP02} or by considering special, experimentally relevant, subclasses such as Gaussian states \cite{EisertP03}.

{\it Conclusions.---}
In this manuscript, we introduced the notion of incoherent states (in a fixed basis) which then allowed us to identify incoherent operations \cite{remark2}. We explicitly distinguished between incoherent operations (A) with and (B) without sub-selection and established the maximally coherent state as the element from which all quantum states (mixed or pure) can be generated only by means of these operations [either type (A) or (B)]. We gave a set of properties which every proper measure of coherence should satisfy and identified the relative entropy of coherence and the $l_1$-norm of coherence as the most general and easy-to-use quantifiers. The questions that we formulated in the outlook of this work are of considerable interest to complete this resource theory of coherence. 

We acknowledge discussions with Susana Huelga that helped to motivate the development of the present work, and discussions with Nathan Killoran and Robert Spekkens.
This work was supported by the Alexander von Humboldt Foundation, the EU Integrating Project SIQS, the EU STREP PAPETS, and the BMBF Verbundprojekt QuoReP.

\vspace{0.5cm}
{\it Note added: While finishing this manuscript, we became aware of the related Ref.
\cite{levi2013aquantitative}, in which questions of monotonicity under incoherent operations
are also discussed.}

%%%%%%%%%%%%%%%%%%%%%%%%%%
% The Appendix
\onecolumngrid
\setcounter{equation}{0}
\renewcommand{\theequation}{A\arabic{equation}}
\numberwithin{equation}{section}

\appendix
\section{\label{sec:TheMaximallyCoherentState}The Maximally Coherent State}
We show that every $d\times d$ state $\hat{\varrho}$ may be prepared from $|\Psi_d\rangle$ by using only incoherent operations.  We do this by
an explicit construction of Kraus operators $\hat{K}_n$, $n=1,\ldots,d$. Let
\begin{equation}
\hat{K}_n = \sum_{i=1}^{d} c_{i} |i\rangle \langle m_{i+n-1}|
\label{eqn:DefinitionMaxCohMap}
\end{equation}
with $c_{i}\in\mathbb{C}$ such that $\sum_{i=1}^{d} |c_{i}|^2 = 1$ and $m_x = \text{mod}(x-1,d)+1 = x - \lfloor \frac{x-1}{d}\rfloor d$. Then
$\sum_{n=1}^{d} \hat{K}_n^{\dagger}\hat{K}_n=\id$ as $\sum_{n}|m_{i+n-1}\rangle \langle m_{i+n-1}| = \sum_{n=1}^d |n\rangle\langle n| = \id$ for all $i$.
Further, for any diagonal density matrix $\hat{\sigma}=\sum_i\sigma_i|i\rangle\langle i|\in\mathcal{I}$, we find
\begin{equation}
\begin{split}
\hat{K}_n\hat{\sigma}\hat{K}_n^\dagger&=\sum_{i,j,k=1}^d
 \sigma_ic_{j}  c_{k}^* \delta_{i,m_{j+n-1}}\delta_{i,m_{k+n-1}}|j\rangle 
 \langle k| \\&= \sum_{k}\sigma_{m_{k+n-1}}|c_{k}|^2|k\rangle\langle k|\in\mathcal{I},
 \end{split}
\end{equation}
i.e., these Kraus operators define an incoherent operation in terms of the two classes of quantum operations (A) and (B), see the main text for further details.
Now, let $|\Psi_d\rangle$ be the maximally coherent state. Then
\begin{equation}
\begin{split}
\hat{K}_n |\Psi_d\rangle &= \frac{1}{\sqrt{d}}\sum_{i=1}^{d} c_{i} |i\rangle \sum_{j=1}^{d} \delta_{j,m_{i+n-1}} \\ &= \frac{1}{\sqrt{d}}\sum_{i=1}^{d} c_{i} |i\rangle.
\end{split}
\end{equation}
Hence, for each outcome $n$, we have
\begin{equation}
\hat{\varrho}_{n} = \frac{\hat{K}_n|\Psi_d\rangle\langle \Psi_d |\hat{K}_n^{\dagger} }{ p_{n}} = |\phi\rangle\langle\phi|
\end{equation}
with probability $p_{n}=1/d$ and where $|\phi\rangle=\sum_{i=1}^{d}c_i|i\rangle$. Thus, with certainty, every pure state $|\phi\rangle$ may be prepared by incoherent operations from the maximally coherent state.
Now let $\hat{\varrho}=\sum_{l}q_l|\phi_l\rangle\langle\phi_l|$, $\sum_lq_l=1$, $|\phi_l\rangle=\sum_ic^{(l)}_i|i\rangle$, be an arbitrary mixed state and define the Kraus operators
\begin{equation}
\hat{K}_n^{(l)}=\sqrt{q_l}\sum_{i=1}^{d} c^{(l)}_{i} |i\rangle \langle m_{i+n-1}|.
\end{equation}
As above, they sum to unity and are incoherent. Further
\begin{equation}
\sum_{n,l}
\hat{K}^{(l)}_n|\Psi_d\rangle\langle \Psi_d |\bigl(\hat{K}^{(l)}_n\bigr)^{\dagger} =\sum_lq_l|\phi_l\rangle\langle\phi_l|=\hat{\varrho},
\end{equation}
i.e., performing generalized measurements according to the $\hat{K}_n^{(l)}$ (and actively erasing the information about measurement outcomes in the case of (B)) prepares the state $\hat{\varrho}$ with certainty.

\section{\label{sec:Gate} Realization of Coherent Gates by Incoherent Operations and Coherent States as Resource}
We set out to show how to implement any unitary operation $\hat{U}=\sum_{i,j=0}^1U_{ij}|i\rangle\langle j|$ only by means of incoherent operations $\{\hat{K}_{i}\}$ and the maximally coherent state
\begin{equation}
|\Psi_2\rangle = \frac{1}{\sqrt{2}} \sum_{l=0}^{1} |l\rangle = \frac{1}{\sqrt{2}}\begin{pmatrix} 1\\1\end{pmatrix}.
\end{equation}
To this end, we let
\begin{equation}
\begin{split}
\hat{K}_{0} = \,&U_{00}|00\rangle\langle00| + U_{10}|10\rangle\langle 01| \\ &+ U_{01}|00\rangle\langle 10| + U_{11}|10\rangle\langle11| 
\end{split}
\end{equation}
and
\begin{equation}
\begin{split}
\hat{K}_{1} = \,&U_{00}|01\rangle\langle01| + U_{10}|11\rangle\langle00|\\ &+ U_{01}|01\rangle\langle11|  +  U_{11}|11\rangle\langle10| 
\end{split}
\end{equation}
be two Kraus operators. Note that (i) $\sum_{i=0}^{1} \hat{K}_{i}^{\dagger}\hat{K}_{i}=\id$ and (ii) $\hat{K}_{i}\mathcal{I}\hat{K}_{i}^{\dagger}\subset\mathcal{I}$ for all $i=0,1$. This can be verified straightforwardly by inspection such that these operators form an incoherent operation as defined in the main text (of type (A) and (B)). Now, let $|\phi\rangle = \sum_{k=0}^{1}\phi_{k}|k\rangle$, and
\begin{equation}
|\xi\rangle = |\phi\rangle\otimes|\Psi_2\rangle = \frac{1}{\sqrt{2}} \sum_{k,l=0}^{1} \phi_{k}|kl\rangle,
\end{equation}
then $\hat{K}_{0} |\xi\rangle= \hat{U}|\phi\rangle\otimes|0\rangle/\sqrt{2}$ and $\hat{K}_{1} |\xi\rangle= \hat{U}|\phi\rangle\otimes |1\rangle/\sqrt{2}$.
Thus, under type (A) and (B) operations, the system will be in the desired state $\hat{U}|\phi\rangle\langle \phi|\hat{U}^\dagger$ with certainty. Recall that this is achieved only by the incoherent operators $\hat{K}_{i}$, $i=0,1$, and the consumption of one maximally coherent state $|\Psi_{2}\rangle$.

\section{\label{sec:FiniteCopyTransformations}Finite Copy Transformations}
In this section, we provide a specific set of Kraus operators that allow us---with finite probability---to transform a pure state into another. For this, let $|\psi\rangle = \sum_{l=1}^{d} \psi_{l} |l\rangle\in\mathbb{C}^{d}$ and $|\phi\rangle = \sum_{l=1}^{d} \phi_{l} |l\rangle\in\mathbb{C}^{d}$ be two pure quantum states. Denote as $M_{\psi}$ and $M_{\phi}$ the number of non-zero coefficients for the respective states, i.e.,
\begin{equation}
M_{\psi} = \Bigl|\bigl\{\psi_{l} \vert \psi_{l}\neq 0 \text{ for }l=1,\ldots,d\bigl\}\Bigl|
\end{equation}
and $M_{\phi}$ equivalently. If $M_{\psi} \geq M_{\phi}$, then one can construct a set of incoherent Kraus operators $\{\hat{K}_{n}\}$ such that
\begin{equation}
\hat{\varrho}_{1} = \frac{\hat{K}_{1}|\psi\rangle \langle \psi|\hat{K}_{1}^{\dagger}}{p_{1}} = |\phi\rangle\langle \phi|
\end{equation}
 with probability $p_{1} = 1 / \sum_{l,\psi_{l}\neq 0} \big\vert \frac{\phi_{l}}{\psi_{l}} \big\vert^2$.

First, assume that $M_{\psi} = d$, i.e., $\psi_{l}\neq 0$ for all $l=1,\ldots,d$. Let
\begin{equation}
\hat{K}_{n} = \sum_{l=1}^{d} \frac{c_{l}}{\psi_{l}} |l\rangle \langle m_{l+n-1}|,
\label{eqn:Akdefinition}
\end{equation}
where $m_{x}$ is defined as above, i.e., $m_x = \text{mod}(x-1,d)+1 = x - \lfloor \frac{x-1}{d}\rfloor d$ and $c_{l} = \phi_{l} \sqrt{p_{1}}$ with $p_{1} = 1 / \sum_{l} \big\vert \frac{\phi_{l}}{\psi_{l}} \big\vert^2$. Note that these operators define incoherent operations as $\hat{K}_{n}\hat{\delta}\hat{K}_{n}^{\dagger}\in\mathcal{I}$ for all $\hat{\delta}=\sum_{i}\delta_{i}|i\rangle\langle i|\in\mathcal{I}$. Further, the Kraus operators satisfy the normalization condition, i.e.,
\begin{equation}
\sum_{n}\hat{K}_{n}^{\dagger}\hat{K}_{n} = \sum_{n}\sum_{l}\bigg\vert\frac{\phi_{l}}{\psi_{l}}\bigg\vert^2 p_{1} |n\rangle\langle n| = \id,
\end{equation}
where, as before, $\sum_{n} |m(l+n-1)\rangle\langle m(l+n-1)| = \sum_{n} |n\rangle\langle n|=\id$ for all $l$. Moreover, we have
\begin{equation}
\hat{K}_1|\psi\rangle = \sum_{k,l=1}^{d}\frac{c_{l}}{\psi_{l}} \psi_{k} |l\rangle\langle l|k\rangle  = \sum_{l=1}^{d} c_{l} |l\rangle = \sqrt{p_1}|\phi\rangle
\end{equation}
such that we find
\begin{equation}
\hat{\varrho}_{1} = \frac{\hat{K}_1|\psi\rangle \langle \psi|\hat{K}_1^{\dagger}}{p_{1}} = |\phi\rangle\langle \phi|
\end{equation}
with probability $p_{1} = \tr\bigl[\hat{K}_1|\psi\rangle\langle\psi |\hat{K}_1^{\dagger}\bigr] = 1 / \sum_{l} \big\vert \frac{\phi_{l}}{\psi_{l}} \big\vert^2$. Now, if $M_{\psi}\neq d$ use the permutations $P_{\psi}$ and $P_{\phi}$ to rearrange the entries of the states such that $P_{\psi}|\psi\rangle = \sum_{l}\psi_{P_{\psi}(l)}|l\rangle$ with $\psi_{P_{\psi}(1)}\geq \ldots\geq \psi_{P_{\psi}(M_{\psi})} > 0$, and similarly for $P_{\phi}|\phi\rangle$. Note that a permutation matrix maps $\mathcal{I}$ onto itself and hence is an incoherent operation. Now, separate the total Hilbert space $\mathcal{H}$  such that $\mathcal{H} = \mathcal{H}_{1} \oplus \mathcal{H}_{2}$ where $\mathcal{H}_{1}$ is of dimension $M_{\psi}$ and $\mathcal{H}_{2}$ is of dimension $d-M_{\psi}$ (note that $P_{\psi}|\psi\rangle$ and $P_{\phi}|\phi\rangle$ are only supported in $\mathcal{H}_{1}$). Apply the Kraus operators $\hat{L}_{n} = \hat{K}_{n} \oplus \mathbb{O}$ for $n=1,\ldots,M_{\psi}$ and $\hat{L}_{n}=\mathbb{O}\oplus |l_{n}\rangle\langle l_{n}|$ with $l_{n} = 1,\ldots,d-M_{\psi}$ for $n=M_{\psi}+1,\ldots,d$ to the state $P_{\psi}|\psi\rangle$. Here, the $\hat{K}_{n}$ are as in
equation \eqref{eqn:Akdefinition} but restricted to the subspace $\mathcal{H}_{1}$ (i.e., the sum is over all $\psi_{l}\neq 0$). Note that the system $\{\hat{L}_n\}$ defines a valid incoherent operation as every element maps diagonal matrices onto diagonal matrices and further satisfies the normalization condition. As before, the application of the operator $\hat{K}_1$ to the subsystem $\mathcal{H}_{1}$ will produce the state $P_{\phi}|\phi\rangle$ in $\mathcal{H}$ with probability $p_{1} = 1 / \sum_{l,\psi_{l}\neq 0} \big\vert \frac{\phi_{l}}{\psi_{l}} \big\vert^2$ where the sum is over all $\psi_{l}\neq 0$ (i.e., in the subspace $\mathcal{H}_{1}$). Applying the inverse of the permutation $P_{\phi}$ (which is an incoherent operation) produces the desired state $|\phi\rangle$ with an overall probability of $p_{1} = 1 / \sum_{l,\psi_{l}\neq 0} \big\vert \frac{\phi_{l}}{\psi_{l}} \big\vert^2$. Note that this protocol may not be optimal, i.e., $P\bigl(|\psi\rangle\mapsto|\phi\rangle\bigl)\geq p_{1}= 1 / \sum_{l,\psi_{l}\neq 0} \big\vert \frac{\phi_{l}}{\psi_{l}} \big\vert^2$, and sub-selection is required (type (B) operations).

\section{\label{sec:L1AndRelEntSatisfyC2c} A third Monotonicity Criterion}

Besides the operations of type (A) and (B) that are discussed in the main text, one may argue that sub-selection based on measurement outcomes is described
by adding a classical flag to the relevant quantum states $\hat{\varrho}_i$, i.e., that one obtains a state of the form $\sum_i p_i |i\rangle\langle i|\otimes\hat{\varrho}_i$, with $p_i$, $\hat{\varrho}_i$ as in (B) and $|i\rangle\langle i|\in \mathcal{I}$. Note that, here, (A) follows by erasing the classical flag, i.e., tracing out the auxiliary system, and (B) may be obtained by projective measurements $\hat{P}_{i}=|i\rangle\langle i|\otimes\id$ (which are incoherent operators with respect to the basis $\{|i\rangle\}$) and tracing over the ancilla. Monotonicity under these incoherent operations would then require:

\begin{enumerate}[leftmargin=1cm]
\item[(C2c)] $C(\hat{\varrho})\ge C\bigl(\sum_i p_i|i\rangle\langle i|\otimes\hat{\varrho}_i\bigr)$ for all $|i\rangle\langle i|\in\mathcal{I}$ and all $\{\hat{K}_i\}$ with $\sum_i\hat{K}_i^\dagger\hat{K}_i=\id$ and $\hat{K}_i\mathcal{I}\hat{K}_i^\dagger\subset\mathcal{I}$.
\end{enumerate}

We find that the relative entropy of coherence and the $l_1$-norm of coherence straightforwardly fulfil this additional constraint as they satisfy (C2b), (C3) and $C(|i\rangle\langle i|\otimes\hat{\varrho}_{i}) \le C(\hat{\varrho}_{i})$:

For the relative entropy, one has
\begin{equation}
\label{sth}
S(\hat{\varrho}\|\hat{\delta})=S(|\alpha\rangle\langle\alpha|\otimes\hat{\varrho}\||\alpha\rangle\langle\alpha|\otimes\hat{\delta})
\end{equation}
for all state vectors $|\alpha\rangle$ and all states $\hat{\varrho}$ and $\hat{\delta}$. Now, if $|i\rangle\langle i|\in\mathcal{I}$ and $\hat{\delta}\in\mathcal{I}$ then
 $|i\rangle\langle i|\otimes\hat{\delta}\in\mathcal{I}$, i.e., whenever $|i\rangle\langle i|\in\mathcal{I}$, we have
\begin{equation}
\begin{split}
&C_{\text{rel.\ ent.}}\Bigl(\sum_ip_i|i\rangle\langle i|\otimes\hat{\varrho}_i\Bigr)\\
&\hspace{1cm}\stackrel{\text{ (C3)}}{\le}
\sum_ip_iC_{\text{rel.\ ent.}}\bigl(|i\rangle\langle i|\otimes\hat{\varrho}_i\bigr)\\
&\hspace{1cm}\le \sum_ip_iS\bigl(|i\rangle\langle i|\otimes\hat{\varrho}_i\||i\rangle\langle i|\otimes\hat{\delta}_i^*\bigr)\\
&\hspace{1cm}\stackrel{\text{\eqref{sth}}}{=}\sum_ip_iS(\hat{\varrho}_i\||\hat{\delta}_i^*)=\sum_ip_iC_{\text{rel.\ ent.}}(\hat{\varrho}_i)\\
&\hspace{1cm}\stackrel{\text{(C2b)}}{\le}C_{\text{rel.\ ent.}}(\hat{\varrho}),
\end{split}
\end{equation}
which is (C2c).

For the $l_1$-norm, we observe that for any $|i\rangle\langle i|\in\mathcal{I}$ and any matrix $\hat{M}$, one has
\begin{equation}
\begin{split}
\bigl\||i\rangle\langle i|\otimes\hat{M}\bigr\|_{l_1}
&
=\sum_{j,k,l,m}
\Bigl|\bigl(|i\rangle\langle i|\otimes\hat{M}\bigr)_{(j,k),(l,m)}\Bigr| \\
&=\sum_{j,k,l,m}
\delta_{j,i}\delta_{l,i}
|M_{k,m}|\\
&=\sum_{k,m}
|M_{k,m}|,
\end{split}
\end{equation}
i.e., $\bigl\||i\rangle\langle i|\otimes(\hat{\varrho}-\hat{\varrho}_{\text{diag}})\bigr\|_{l_1}= \bigl\|\hat{\varrho}-\hat{\varrho}_{\text{diag}}\bigr\|_{l_1}$, and therefore, as above,
\begin{equation}
\begin{split}
&C_{l_1}\Bigl(\sum_ip_i|i\rangle\langle i|\otimes\hat{\varrho}_i\Bigr)\\
&\hspace{1cm}\stackrel{\text{(C3)}}{\le}
\sum_ip_iC_{l_1}\bigl(|i\rangle\langle i|\otimes\hat{\varrho}_i\bigr)\\
&\hspace{1cm}\le \sum_ip_i\bigl\||i\rangle\langle i|\otimes\hat{\varrho}_i-|i\rangle\langle i|\otimes\hat{\varrho}_i^{\text{diag}}\bigr\|_{l_1}\\
&\hspace{1cm}=\sum_ip_i\bigl\|\hat{\varrho}_i-\hat{\varrho}_i^{\text{diag}}\bigr\|_{l_1}=\sum_ip_iC_{l_1}(\hat{\varrho}_i)\\
&\hspace{1cm}\stackrel{\text{(C2b)}}{\le}C_{l_1}(\hat{\varrho}),
\end{split}
\end{equation}
which is condition (C2c).

\section{\label{sec:RelativeEntropyOfCoherence} {(C2\lowercase{b})} for the Relative Entropy of Coherence}
We set out to establish the monotonicity criterion for the relative entropy of coherence $C_{\text{rel.\ ent.}}=\min_{\delta\in\mathcal{I}}S(\hat{\varrho}\|\hat{\delta})$ for condition (C2b), i.e., we show that 
\begin{equation}
C_{\text{rel.\ ent.}}(\hat{\varrho}) \geq \sum_np_nC_{\text{rel.\ ent.}}(\hat{\varrho}_n)
\end{equation}
for all $\{\hat{K}_n\}$ with $\sum_n\hat{K}_n^\dagger\hat{K}_n=\id$ and $\hat{K}_n\mathcal{I}\hat{K}_n^\dagger\subset\mathcal{I}$. Let $\hat{\varrho}_{n} = \hat{K}_{n}\hat{\varrho}\hat{K}_{n}^{\dagger}/p_{n}$ with $p_{n} = \text{tr}[\hat{K}_{n}\hat{\varrho}\hat{K}_{n}^{\dagger}]$, then
\begin{equation}
S(\hat{\varrho}\|\hat{\delta})\ge \sum_np_nS(\hat{\varrho}_n\|\hat{K}_n\hat{\delta}\hat{K}_n^\dagger/\text{tr}[\hat{K}_n\hat{\delta}\hat{K}_n^\dagger]).
\end{equation}
This follows as the quantum relative entropy satisfies conditions (F1)--(F5) in \cite{vedral1998entanglement}. With this, we have
\begin{equation}
\begin{split}
C_{\text{rel.\ ent.}}(\hat{\varrho})&=S(\hat{\varrho}\|\hat{\delta}^*) \\
&\ge \sum_np_nS(\hat{\varrho}_n\|\hat{K}_n\hat{\delta}^*\hat{K}_n^\dagger/\text{tr}[\hat{K}_n\hat{\delta}^*\hat{K}_n^\dagger])\\
&\ge \sum_np_n\min_{\delta\in\mathcal{I}}S(\hat{\varrho}_n\|\hat{\delta})\\
&=\sum_np_nC_{\text{rel.\ ent.}}(\hat{\varrho}_n),
\end{split}
\end{equation}
as $\hat{K}_{n}\hat{\delta}^{*}\hat{K}_{n}^{\dagger}\in\mathcal{I}$. This proves (C2b) for the quantum relative entropy.

\newcommand{\lone}{l_1}
\section{\label{sec:l1NorOfCoherence} (C2\lowercase{b}) for the \boldmath{$\protect\lone$}-Norm of Coherence}
We show monotonicity for the $l_1$-norm of coherence according to (C2b). Recall the closed form of this coherence measure, that is, 
\begin{equation}
C_{l_1}(\hat{\varrho})=\sum_{\substack{i,j\\ i\ne j}}|\varrho_{i,j}|. 
\end{equation}
Now, for given $\hat{\varrho}$, consider
\begin{equation}
\begin{split}
\sum_np_n
C_{l_1}(\hat{\varrho}_n)&=\sum_np_n\sum_{\substack{i,j\\ i\ne j}}|[\hat{\varrho}_n]_{i,j}| \\
&=\sum_n
\sum_{\substack{i,j\\ i\ne j}}|[\hat{K}_n\hat{\varrho}\hat{K}_n^\dagger]_{i,j}|\\
&=\sum_n\sum_{\substack{i,j\\ i\ne j}}\Bigl|\sum_{k,l}[\hat{K}_n]_{i,k}\varrho_{k,l}[\hat{K}_n^\dagger]_{l,j}\Bigr|,
\end{split}
\end{equation}
where, denoting by $\hat{\varrho}_{\text{diag}}$ the incoherent state $\hat{\varrho}_{\text{diag}}=\sum_k\varrho_{k,k}|k\rangle\langle k|$, we have 
\begin{equation}
\begin{split}
\sum_{k}[\hat{K}_n]_{i,k}\varrho_{k,k}[\hat{K}_n^\dagger]_{k,j}&=\bigl(\hat{K}_n\hat{\varrho}_{\text{diag}}\hat{K}_n^\dagger\bigr)_{i,j}\\
&=\delta_{i,j}\bigl(\hat{K}_n\hat{\varrho}_{\text{diag}}\hat{K}_n^\dagger\bigr)_{i,i},
\end{split}
\end{equation}
i.e.,
\begin{equation}
\begin{split}
\sum_np_n
C_{l_1}(\hat{\varrho}_n)&=\sum_n\sum_{\substack{i,j\\ i\ne j}}\Bigl|\sum_{\substack{k,l\\ k\ne l}}[\hat{K}_n]_{i,k}\varrho_{k,l}[\hat{K}_n^\dagger]_{l,j}\Bigr| \\
&\le \sum_{\substack{k,l\\ k\ne l}}|\varrho_{k,l}|
\sum_n\sum_{\substack{i,j\\ i\ne j}}|[\hat{K}_n]_{i,k}[\hat{K}_n^\dagger]_{l,j}|.
\end{split}
\end{equation}
Further, we find
\begin{equation}
\begin{split}
&\sum_n\sum_{\substack{i,j\\ i\ne j}}|[\hat{K}_n]_{i,k}[\hat{K}_n^\dagger]_{l,j}| \\
&\hspace{0.5cm}\le
\sum_n\sum_i|[\hat{K}_n]_{i,k}|\sum_j |[\hat{K}_n^\dagger]_{l,j}| \\
&\hspace{0.5cm}\le \sqrt{\sum_n\Bigl(\sum_i|[\hat{K}_n]_{i,k}|\Bigr)^2\sum_m\Bigl(\sum_j |[\hat{K}_m^\dagger]_{l,j}|\Bigr)^2}
\end{split}
\end{equation}
and
\begin{equation}
\begin{split}
&\sum_n\Bigl(\sum_i|[\hat{K}_n]_{i,k}|\Bigr)^2\\
&\hspace{1cm}=\sum_n\sum_{i,j}|[\hat{K}_n]_{i,k}[\hat{K}^\dagger_n]_{k,j}| \\
&\hspace{1cm}=
\sum_n\sum_{i,j}|\langle i|\hat{K}_n|k\rangle\langle k|\hat{K}^\dagger_n|j\rangle|\\
&\hspace{1cm}=
\sum_n\sum_{i,j}\delta_{i,j}|\langle i|\hat{K}_n|k\rangle\langle k|\hat{K}^\dagger_n|j\rangle|\\
&\hspace{1cm}=\sum_n\sum_{i}|\langle i|\hat{K}_n|k\rangle\langle k|\hat{K}^\dagger_n|i\rangle|\\
&\hspace{1cm}=\sum_n\sum_{i}\langle k|\hat{K}^\dagger_n|i\rangle\langle i|\hat{K}_n|k\rangle=1,
\end{split}
\end{equation}
such that
\begin{equation}
\begin{split}
\sum_np_n
C_{l_1}(\hat{\varrho}_n)&\le \sum_{\substack{k,l\\ k\ne l}}|\varrho_{k,l}|=C_{l_1}(\hat{\varrho}).
\end{split}
\end{equation}
This proves (C2b) for the $l_1$-norm of coherence. Furthermore, this---together with the convexity of the measure, i.e., condition (C3)---implies monotonicity under (C2a) as discussed in the main text. 

\section{\label{sec:HSNormIsNoMonotone} Violation of (C2\lowercase{b})}
In this section of the Appendix, we show that
\begin{equation}
C_{l_2}(\hat{\varrho}):= \min_{\hat{\sigma}\in\mathcal{I}} \|\hat{\varrho}-\hat{\sigma}\|_{l_{2}}^{2}=\sum_{\substack{i,j\\ i\ne j}}|\varrho_{i,j}|^2
\end{equation}
does not satisfy monotonicity under (C2b). We prove this fact by establishing a counter-example.
For this, we construct a set of Kraus operators together with a specific quantum state and show that the functional $C_{l_2}$ can increase on average under incoherent operations. Let
\begin{equation}
\hat{K}_{1} = \begin{pmatrix}  0 & 1 & 0 \\ 0 & 0 & 0 \\ 0 & 0 & \alpha \end{pmatrix}
\; \text{ and } \;
\hat{K}_{2} = \begin{pmatrix} 1 & 0 & 0 \\ 0 & 0 & \beta \\ 0 & 0 & 0 \end{pmatrix}
\end{equation}
with $\alpha, \beta\in\mathbb{C}$ and $|\alpha|^2 + |\beta|^2 = 1$. Note that the latter constraint guarantees that $\sum_{n} \hat{K}_{n}^{\dagger}\hat{K}_{n}=\id$. Moreover, we have $\hat{K}_{n}\hat{\delta}\hat{K}^{\dagger}_{n}\in\mathcal{I}$ for $n=1,2$ and all $\hat{\delta}\in\mathcal{I}$. Hence, in none of the outcomes the quantum operations $\{\hat{K}_{n}\}$ generate coherence from incoherent states. Now, consider the state
\begin{equation}
\hat{\varrho} = \frac{1}{2} \bigl[|\psi_{1}\rangle\langle\psi_{1}| + |\psi_{2}\rangle\langle\psi_{2}|\bigr]
\end{equation}
with $|\psi_{1}\rangle = [0\;1\;0]^{\text{T}}$ and $|\psi_{2}\rangle = [1\;0\;1]^{\text{T}}/\sqrt{2}$. We find $C_{l_2}(\hat{\varrho}) = 1/8$ and
\begin{equation}
\sum_{k=1}^{2}p_{k} C_{l_2}(\hat{\varrho}_{k}) = p_{2} C_{l_2}(\hat{\varrho}_{2}) = \frac{|\beta|^2}{2(1+|\beta|^2)}.
\end{equation}
For $|\beta |^2 > 1/3$ monotonicity in formulation (C2b) is violated, that is, $\sum_{k=1}^{2}p_{k} C_{l_2}(\hat{\varrho}_{k})>C_{l_2}(\hat{\varrho})$. This can be achieved with, e.g., $\alpha = 1/\sqrt{2} = \beta$. In words, allowing for sub-selection, the sum of the absolute values squared of the off-diagonal elements is not a proper measure to quantify coherence in a quantum system.

\section{\label{sec:OtherCandidates} Other Candidates for Coherence Measures}
In the last section of the Appendix, we briefly comment on other possible candidates for coherence measures. Consider the fidelity between quantum states \cite{Uhlmann1976}
\begin{equation}
F(\hat{\varrho},\hat{\delta})=\text{tr}\left[\sqrt{\hat{\varrho}^{1/2}\hat{\delta}\hat{\varrho}^{1/2}}\right]^2=\|\hat{\varrho}^{1/2}\hat{\delta}^{1/2}\|_{\text{tr}}^2.
\end{equation}
It is known that $\sqrt{F}$ is jointly concave, non-decreasing under CPTP maps, and $F(\hat{\varrho},\hat{\delta})=1$ iff $\hat{\varrho}=\hat{\delta}$, see, e.g., Refs.~\cite{Barnum1996,Miszczak2009} and references therein. Hence, the coherence measure induced by
\begin{equation}
\mathcal{D}(\hat{\varrho},\hat{\delta})=1-\sqrt{F(\hat{\varrho},\hat{\delta})}
\end{equation}
fulfils (C1'), (C2a), and (C3).

The trace norm 
\begin{equation}
\mathcal{D}_{\text{tr}}(\hat{\varrho},\hat{\delta})=\|\hat{\varrho}-\hat{\delta}\|_{\text{tr}}
\end{equation}
is a matrix norm and contracting under CPTP maps \cite{Wolf_contractivity}, i.e., as discussed in the main text, the induced measure of coherence fulfils (C1'), (C2a), and (C3). Note, however, that in general $\min_{\hat{\delta}\in\mathcal{I}} \|\hat{\varrho}-\hat{\delta}\|_{\text{tr}}\ne \|\hat{\varrho}-\hat{\varrho}_{\text{diag}}\|_{\text{tr}}$.


\begin{thebibliography}{00}
\bibitem{remark5}
Note that we do not specify the Hilbert space structure. In particular, $\mathcal{H}$ may describe compound quantum systems, e.g., $\mathcal{H} = \mathbbm{C}^{2}\otimes\ldots\otimes \mathbbm{C}^{2}$ for a set of qubits.
\bibitem{remark4}
To simplify notation, we do not specify the dimension $d$, the context should make this unambiguous.
\bibitem{remark1}
Note the close relation to $U(1)$-covariant operations considered in the resource theory of reference frames \cite{GourS08}.
These operations are incoherent in the sense that they map diagonal states to diagonal states, they are, however, a smaller set of operations,
in particular, they do not allow for certain permutations.
\end{thebibliography}
\end{document}